# Liquid-liquid transition in supercooled aqueous solution involving a low-temperature phase similar to low-density amorphous water

Sander Woutersen[*], Michiel Hilbers[*], Zuofeng Zhao[#] and C. Austen Angell[#]

[*]Van 't Hoff Institute for Molecular Sciences, University of Amsterdam, Science Park 904,1098 XH Amsterdam, The Netherlands

[#]School of Molecular Sciences, Arizona State University, Tempe, AZ, 85287-1604, USA

**The striking anomalies in physical properties of supercooled water that were discovered in the 1960-70s, remain incompletely understood and so provide both a source of controversy amongst theoreticians[1-5], and a stimulus to experimentalists and simulators to find new ways of penetrating the "crystallization curtain" that effectively shields the problem from solution[6,7]. Recently a new door on the problem was opened by showing that, in ideal solutions, made using ionic liquid solutes, water anomalies are not destroyed as earlier found for common salt and most molecular solutes, but instead are enhanced to the point of precipitating an apparently first order liquid-liquid transition (LLT)[8]. The evidence was a spike in apparent heat capacity during cooling that could be fully reversed during reheating before any sign of ice crystallization appeared. Here, we use decoupled-oscillator infrared spectroscopy to define the structural character of this phenomenon using similar down and upscan rates as in the calorimetric study. Thin-film samples also permit slow scans (1 Kmin$^{-1}$) in which the transition has a width of less than 1 K, and is fully reversible. The OH spectrum changes discontinuously at the phase-transition temperature, indicating a discrete change in hydrogen-bond structure. The spectral changes show that the low-temperature liquid is more strongly hydrogen bonded and less disordered as compared to the high-temperature liquid. The spectrum of the low-temperature liquid is essentially that seen in low-density amorphous (LDA) water. This similarity suggests that the liquid-liquid transition observed here also exists in neat undercooled water, providing a unified explanation for many of its anomalies.**

      The notion that studies of glass-forming aqueous solutions can provide evidence of liquid-liquid phase separation actually precedes the findings of anomalies in supercooled pure water. Initial conjectures[9] were based on differential thermal analysis (DTA) studies of the lithium chloride + H$_2$O system, and an analogy was drawn with the behavior of the Na$_2$O-SiO$_2$ system, in ignorance of the fact that the second "liquid" phase in the latter system is almost pure SiO$_2$ and thus a glassy phase at the temperature of observed phase separation.[10] Mishima[11] subsequently refined the studies of LiCl-H$_2$O and found a domain at low temperatures where pure amorphous water separated from the saline solution during cooling, but then crystallized to ice Ih with moderate increases in temperature.

      In the meantime, the divergent behavior of pure water thermodynamic properties (and also relaxation times) during deep supercooling had been revealed and then famously interpreted by Poole *et al.*[12] in terms of the existence of a nearby second critical point at which two liquid phases differing only in density, become identical. A

thermodynamic model–a variant of the van der Waals equation of state, was soon devised by Poole[13], to show how the two critical points could be related through the splitting of the familiar van der Waals coexistence domain into two segments. This can happen by the intercession of a low free energy liquid structure, with a density that is somewhere between gas and close-packed liquid, consequent on the existence of a strongly interacting low coordination number network of hydrogen bonds, as in ice and in random tetrahedral network models of water (e.g. Sceats and Rice[14]).

The problem is that, for laboratory water, none of this latter scenario can be established by direct experimentation because of the pre-empting crystallization of ice Ih which forms by homogeneous nucleation and growth in a pattern that closely follows the pattern of divergencies[6], i.e. critical nuclei are a part of the spectrum of entropy fluctuations that determine the heat capacity, or of density fluctuations that determine the compressibility. Increase of pressure causes a decrease in the nucleation temperature but also changes the water structure so as to decrease the divergence temperature[6]. Eventually a new phase of ice becomes preferred before crystallization rates fall low enough to open up the additional temperature range (down to vitrification) that might reveal the liquid-liquid transition.

The addition of second components can serve to depress crystallization rates, and permit vitrification (thus acting as a proxy for pressure increases, as has long been recognized). Theory[15-17] has indicated the possibility of obtaining information on hidden critical phenomena in single component systems by virtue of critical lines emanating from the pure solvent into the binary solution. Unfortunately, most ionic[18,19] and many molecular[20] second components destroy the water anomalies more rapidly than they block the nucleation of ice. Recently, however, a class of ionic solutes has been discovered for which the latter discouraging scenario is reversed, i.e. they permit supercooling to the point of vitrification without destruction of the liquid state anomalies, indeed they seem to enhance them.[8] A consequence is that finally the crystallization curtain can be lifted. These solutions are thermodynamically ideal according to melting point depression data (see Supporting Information of ref. 8).

Figure 1 shows the behavior that is then revealed. Instead of salt addition leading to a replacement of the pure water anomaly with a decreasing solution $C_p$ and finally glass transition (as shown for the familiar case of an 11.4 mol% LiCl solution[19]) we see a first order-like spike in the heat capacity in the case of the solutions of ionic liquid hydrazinium trifluoroacetate ($N_2H_5$.TFA) - though the salt content is higher for the latter case. The "spike anomaly" was fully reversible on reheating. It was followed, at higher temperature, by crystallization of ice $I_h$. The temperature of re-dissolution of the last ice crystal during further heating permitted the liquidus temperature to be determined and it was shown that the solutions were obeying the laws of ideal solutions for fully dissociating monovalent salt solutions.[21]

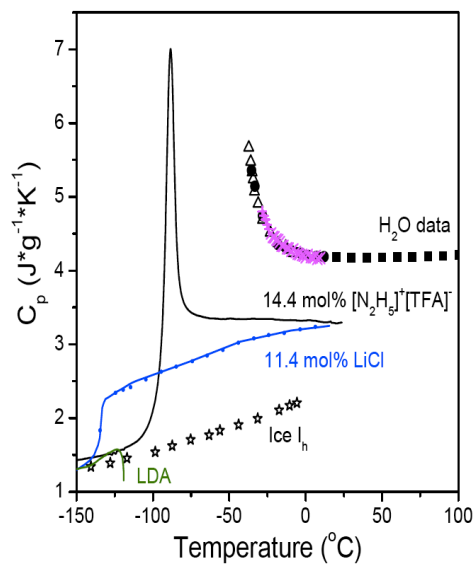

**Figure 1.** Contrasting thermal behavior of ideal and non-ideal aqueous solutions during supercooling. Compositions are identified on the plots. TFA is trifluoroacetate. (Apparent) heat capacities were determined during cooling scans at 20 Kmin$^{-1}$ in each case. Pure H$_2$O data are a compendium of literature results (see ref. 8). Data for LiCl solution were taken from ref. 19.

In ref. 8 several examples of this behavior were shown, but that displayed in Figure 1 is the most pronounced. It occurs in the solution with the highest water content that could be studied without ice crystallization interfering. A related phenomenon has also been reported by Murata and Tanaka[22] using glycerol as the solute, but in this case the formation of the second liquid phase was accompanied by ice formation and so was irreversible. It has been controversial[23] though the sequential nature of liquid-liquid and ice crystallization events has been clarified by 2D-IR spectroscopy.[24] The question now arises as to what are the participating structures in the new first order transition, and it is this question that we address in this communication. As a probe, the decoupled vibration frequency of the OH oscillator is ideal, since the water molecules are the only species possessing this vibration.

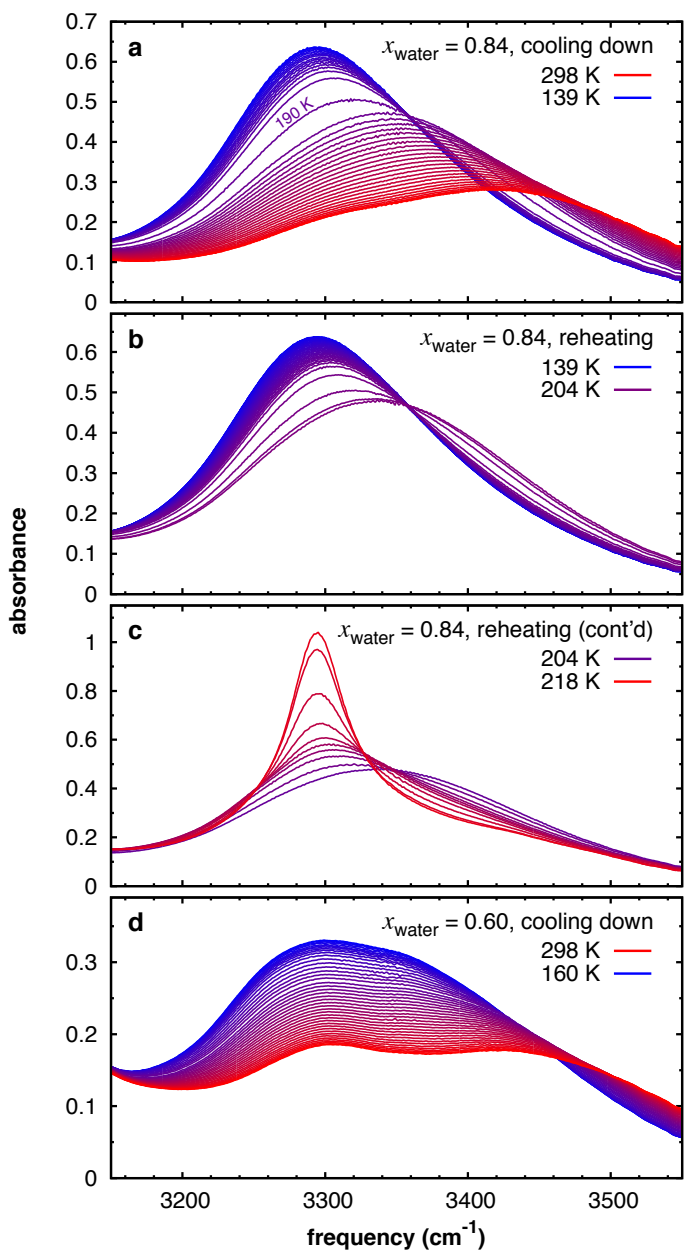

**Figure 2**. (a) IR absorption spectra of a solution of ($N_2HD_4$/$N_2D_5$).TFA in HDO/$D_2O$ (H/D fraction 3%), with a molar water fraction $x = 0.84$, and a sample-cell thickness of 23 μm. Upon decreasing the temperature (at 7 K min$^{-1}$), the OH-stretch peak shows a gradual redshift. At a temperature of 190 K an intense, broad low-frequency OH-stretch mode appears, indicating a discrete change in hydrogen-bond structure. (b) Upon reheating, this phenomenon is reversed, and the low-frequency peak disappears again. (c) Upon further reheating, the water crystallizes, resulting in a narrow low-frequency OH-stretch peak (due to ice) superimposed on a broad OH-stretch mode due to the surrounding liquid. (d) In a solution with a water fraction $x = 0.60$, no structural transition occurs, but only a gradual lowering of the OH-stretch frequency with decreasing temperature.

We prepare aqueous solutions of hydrazinium trifluoroacetate ($N_2H_5$.TFA) with molar water fractions $x_{water}$ ranging from 0.50 to 0.84. The sample volume is a ~1 μl droplet, kept between two $CaF_2$ windows separated by 23 μm using a circular spacer. The small sample volume allows for comparatively slow temperature scans without the occurrence of crystallization. We use dilute H/D isotopic mixtures (with an H:D fraction of 0.03) to prevent coupling between the OH (and NH) stretch modes.[25,26] This ensures that the center frequency and width of the observed OH-stretch absorption peak reflect the hydrogen-bond structure.[27] Fig. 2a shows the IR absorption spectrum of an $x_{water}$ = 0.84 solution at room temperature. The OH/NH-stretch region contains a broad intense band centered at 3420 $cm^{-1}$, and a much weaker peak at 3300 $cm^{-1}$. The latter is absent in the spectrum of an NaTFA solution with the same $x_{water}$ (see Extended Data), which shows that the 3420 $cm^{-1}$ peak is due to the OH-stretch mode of water (HDO), and the 3300 $cm^{-1}$ peak to the NH-stretch mode of hydrazinium ($N_2HD_4^+$). This assignment is confirmed by the change in relative intensities of the two peaks upon changing the water fraction (Fig. 2d). The OH-stretch mode of the solution has a center frequency and width (245 $cm^{-1}$ FWHM) similar to those of liquid HDO:$D_2O$,[28] and in both cases the line shapes are Gaussian. These similarities indicate that even at the high concentration of ions in the solution, the distribution of hydrogen-bond strengths is similar to that of neat water, as is reflected at the macroscopic level by the ideal-solution behavior already mentioned.

Upon cooling down the sample, the frequency of the OH-stretch frequency initially decreases gradually, in a similar fashion as neat undercooled water[29]. But at the temperature of the heat-capacity spike (~190 K), a discontinuous change occurs: a new OH-stretch mode appears at 3300 $cm^{-1}$, and the initial OH-stretch mode vanishes. Upon further cooling, this new OH-stretch mode also gradually decreases in frequency. The transition is reversible (Fig. 2b): upon reheating the sample, the low-frequency OH-stretch mode disappears, and the original high-frequency OH-stretch mode reappears. The sample remains transparent during cooling, and during subsequent reheating until a temperature of about 204 K, see Figure 3. At that point, ice crystallites start to form (Fig. 3, rightmost panel). The presence of ice shows up distinctly in the IR spectrum (Fig. 2c) as a narrow peak at 3295 $cm^{-1}$ with a width of 50 $cm^{-1}$. These numbers agree well with the center frequency and width of the OH-stretch mode of neat HDO:$D_2O$ ice at this temperature.[28] The complete absence of the ice peak in the earlier stages of the temperature scan demonstrates that no ice is formed during the structural transition associated with the heat-capacity spike, neither during cooling nor during reheating. This is in contrast to glycerol:water mixtures,[22] for which IR experiments show evidence that (nanoscopic) ice crystals are formed during or immediately after the putative liquid-liquid transition.[24]

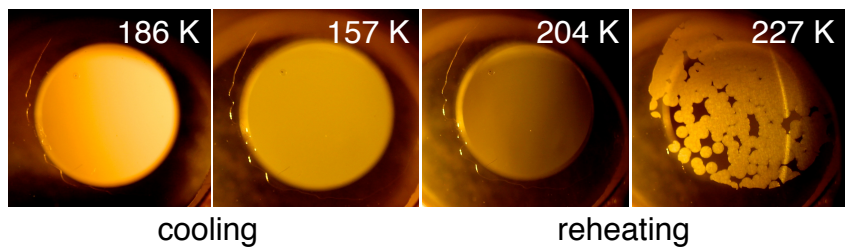

**Figure 3.** (a) Droplet of $(N_2HD_4/N_2D_5)$.TFA solution ($x_{water}$ = 0.84) squashed between two $CaF_2$ windows, during cooling and subsequent reheating. The blurred circle is the 5 mm diameter optical exit of the sample holder, around which the circumference of the droplet is visible as a sharp squiggly line. During reheating, crystallization occurs (rightmost panel), but only after the LLT.

The discontinuous nature of the observed structural transition is clearly visible when monitoring the absorbance at 3300 cm$^{-1}$ during cooling and reheating (Fig. 4). When cooling at a rate of 7 K min$^{-1}$ (red points), the appearance (during cooling) and disappearance (during reheating) of the 3300 cm$^{-1}$ peak occur at temperatures which differ by about 10 K. However, when cooling at a rate of 1 K min$^{-1}$ (blue points in Fig. 4) the difference is only 3 K; and this small difference probably originates mainly from the sample temperature lagging behind the temperature of the thermo-couple, which is not immersed in the liquid but attached to the brass holder keeping the $CaF_2$ windows together (see Extended Data).

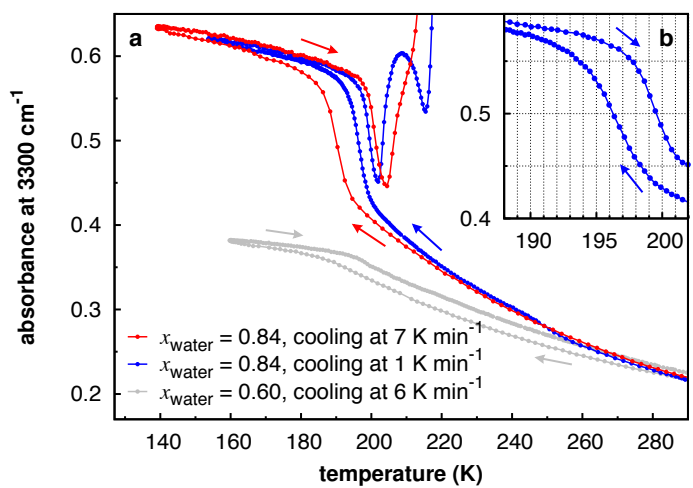

**Figure 4.** (a) IR absorbance at 3300 cm$^{-1}$ versus temperature, for three experiments in which the samples of Fig. 1 were cooled down and then reheated, as indicated by the arrows. Each point indicates a measured IR spectrum. The $x$ = 0.84 sample exhibits discrete transition, which is reversible. At lower cooling and reheating rate the hysteresis is much less. (b) Close up showing the reduced hysteresis at lower cooling rate. Assuming identical temperature lags for cooling and reheating we estimate the transition temperature to be 198K.

The discontinuous change in the OH-stretch spectrum indicates that heat-capacity spike involves an abrupt change in the hydrogen-bond structure. This transition can therefore not be a glass transition, which is predominantly an increase in relaxation time rather than in structure. The difference is nicely illustrated by a measurement on a sample with a water fraction of 0.60, which exhibits a glass transition, visible in the DSC scan as a step decrease in the heat capacity (like in Fig. 1 for the LiCl solution but at higher temperature); in the IR spectrum however (Fig. 2d, and grey points in Fig. 4) only a gradual shift and no discontinuity of the OH-stretch mode is observed (note that the $x_{water}$ = 0.84 and $x_{water}$ = 0.60 data were obtained with the same cooling rate).

Based on the IR spectra it can thus be excluded that the heat-capacity spike is due to ice formation or to a glass transition. The observed change in hydrogen-bond structure rather indicates that the heat-capacity spike is associated with the liquid-liquid phase transition predicted for water by Poole and many others. The OH-stretch spectrum provides direct information on the differences in hydrogen-bond structures between the high-temperature and low-temperature liquid phases (Fig. 5): the lower center frequency implies stronger (shorter) hydrogen bonds in the low-temperature liquid (as compared to the high-temperature liquid), whereas the smaller width (~180 cm$^{-1}$ FWHM) indicates that the hydrogen-bond structure is more ordered. The spectrum of the low-temperature liquid is similar to that of low-density amorphous (LDA) water (dashed curve in Fig. 5, data kindly provided by the authors of Ref. 30). It has a somewhat larger width, indicating more disorder, which is probably due to the presence of the ions. The similarity of these spectra strongly suggests that the liquid-liquid transition observed here is directly connected to the HDA-LDA transition in water; and based on the equivalence of high ionic concentrations with the application of high pressure this connection would imply that a liquid-liquid transition also exists in neat water, providing an explanation for many of its well-known anomalous properties.

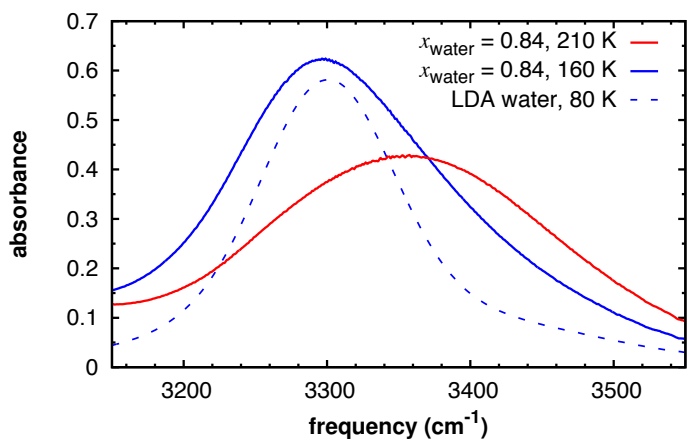

**Figure 5.** IR absorption spectra of the high-temperature phase, the low-temperature phase, and of low-density amorphous water (data from Ref. 30, kindly provided by the authors). All samples are dilute isotopic H:D mixtures.

**References**

1    Liu, Y., Panagiotopoulos, A. Z. & Debenedetti, P. G. Low-temperature fluid-phase behavior of ST2 water. *J. Chem. Phys*. **131**, 104508 (2009).
2    Kesselring, T. A. *et al*. Finite-size scaling investigation of the liquid-liquid critical point in ST2 water and its stability with respect to crystallization. *J. Chem. Phys*. **138**, 244506 (2013).
3    Limmer, D. T. & Chandler, D. The Putative Liquid-Liquid Transition is a Liquid-Solid Transition in Atomistic Models of Water. *J. Chem. Phys*. **135**, 134503 (2011).
4    Limmer, D. T. & Chandler, D. The putative liquid-liquid transition is a liquid-solid transition in atomistic models of water. II. *J. Chem. Phys*. **138,**, 214504; doi: 214510.211063/214501.4807479 (2013).
5    Palmer, J. C. *et al*. Metastable liquid-liquid transition in a molecular model of water. *Nature* **510**, 385 doi:310.1038/nature13405 (2014).
6    Kanno, H. & Angell, C. A. Water: anomalous compressibilities to 1.9 kbar and correlation with supercooling limits. *J. Chem. Phys,* **70**, 4008-4016 (1979).
7    Sellberg, J. A. *et al*. Ultrafast X-ray probing of water structure below the homogeneous ice nucleation temperature. *Nature* **510**, 381-385 (2014).
8    Zhao, Z.-F. & Angell, C. A. Apparent first order liquid-liquid transition with pre-transition density anomaly, in water-rich ideal solutions. *Angewandte Chemie Int*. *Ed*. **55**, 2474-2477 (2016).
9    Angell, C. A. & Sare, E. J. Liquid-Liquid Immiscibility in Common Aqueous Salt Solutions at Low Temperatures,. *J. Chem. Phys*. **49**, 4713 (1968).
10   Doremus, R. *Glass Science*. (Wiley, 1973).

11    Mishima, O. Phase separation in dilute LiCl–H$_2$O solution related to the polyamorphism of liquid water. *J. Chem. Phys.* **126**, 244507 (2007).
12    Poole, P. H., Sciortino, F., Essmann, U. & Stanley, H. E. Phase-Behavior of Metastable Water. *Nature* **360**, 324-328 (1992).
13    Poole, P. H., Sciortino, F., Grande, T., Stanley, H. E. & Angell, C. A. Effect of Hydrogen Bonds on the Thermodynamic Behavior of Liquid Water. *Phys. Rev. Lett.* **73**, 1632-1635 (1994).
14    Sceats, M. G. & Rice, S. A. *Amorphous Solid Water*. Vol. 7 (Plenum Press, 1982).
15    Chatterjee, S. & Debenedetti, P. G. Fluid-Phase Behavior of Binary Mixtures in which One Component Can Have Two Critical Points. *J. Chem. Phys.* **124**, 154503 (2006).
16    Anisimov, M. A. Cold and supercooled water: A novel supercritical-fluid solvent. *Russ. J. Phys. Chem. B* **6**, 861-867 (2012).
17    Biddle, J. W., Holten, V. & Anisimov, M. A. Behavior of supercooled aqueous solutions stemming from hidden liquid–liquid transition in water. *J. Chem. Phys.* **141**, 074504 (2014).
18    Archer, D. G. & Carter, R. W. Thermodynamic properties of the NaCl+H$_2$O system. 4. Heat capacities of H$_2$O and NaCl(aq) in cold-stable and supercooled states *J. Phys. Chem. B* **104**, 8563-8584 (2000).
19    Angell, C. A. Insights into Phases of Liquid Water from Study of Its Unusual Glass-Forming Properties *Science* **319**, 582-587 (2008).
20    Oguni, M. & Angell, C. A. Heat capacities of H$_2$O + H$_2$O$_2$, and H$_2$O + N$_2$H$_4$, binary solutions: Isolation of a singular component for C$_p$ of supercooled water. *J. Chem. Phys.* **73**, 1948-1954 (1980).
21    Moore, W. J. *Physical Chemistry, 4th Edition, 1972,Prentice-Hall, N.J, Chapter 7, section 13*.
22    Murata, K.-I. & Tanaka, H. Liquid–liquid transition without macroscopic phase separation in a water–glycerol mixture. *Nature Materials* **11**, 436-442 (2012).
23    Suzuki, Y. & Mishima, O. Experimentally proven liquid-liquid critical point of dilute glycerol-water solution at 150K. *J. Chem. Phys.* **141**, 094505 (2014).
24    Bruijn, J. R., van der Loop, T. & Woutersen, S. Changes in hydrogen-bond structure during an aqueous liquid-liquid transition investigated with time-resolved and two-dimensional vibrational spectroscopy. *J. Phys. Chem. Lett. (in press)* (2016).
25    Woutersen, S. & Bakker, H. J. Resonant intermolecular transfer of vibrational energy in liquid water. *Nature* **403**, 507-509 (1999).
26    Yang, M. & Skinner, J. L. Signatures of coherent vibrational energy transfer in IR and Raman line shapes for liquid water. *Phys. Chem. Chem. Phys.* **12**, 982-991 (2010).
27    Fecko, C. J., Eaves, J. D., Loparo, J. J., Tokmakoff, A. & Geissler, P. L. Ultrafast hydrogen-bond dynamics in the infrared spectroscopy of water. *Science* **301**, 1698-1702 (2003).
28    Ford, T. A. & Falk, M. Hydrogen bonding in water and ice. *Can. J. Chem.* **46**, 3579-3586 (1968).
29    Perakis, F. & Hamm, P. Two-Dimensional Infrared Spectroscopy of Supercooled Water. *J. Phys. Chem. B* **115**, 5289-5293 (2011).
30    Shalit, A., Perakis, F. & Hamm, P. Two-Dimensional Infrared Spectroscopy of Isotope-Diluted Low Density Amorphous Ice. *J. Phys. Chem. B* **117**, 15512-15518 (2013).

**Acknowledgements**
The authors acknowledge Hans Sanders for preparing the samples. SW acknowledges financial support from the John van Geuns foundation. CAA has been assisted by NSF under Grant no. CHE 12-13265.

**Methods**

*Sample preparation*. A typical preparation would be as follows: a 10 ml reagent tube with NS14/23 connection is cleaned, rinsed and dried by heating with a heat gun and subsequently cooled with a stream of dry nitrogen. The tube is filled with argon, stoppered and tared. Ca. 0.2 ml of N$_2$D$_4$.D$_2$O is added, the tube stoppered and the exact weight of the hydrazine (no density known) is determined. The calculated amount of TFA-d (d=1.493) is then added slowly at 0 °C under a stream of dry nitrogen. The required dilution is obtained by adding the appropriate amount of D$_2$O (and optionally H$_2$O), taking into account the D$_2$O already present from the hydrazine. The samples are then stored and handled under argon. We squash a droplet of the appropriate volume between two 2 mm thick a CaF$_2$ windows separated by a 23 micron spacer, making sure that the liquid is not in contact with the spacer.

*Cryogenic FTIR measurements*. The sample is mounted in a brass holder in thermal contact with the heat exchanger of a liquid-$N_2$ cryostat (DN1704, Oxford Instruments). To monitor the temperature a type K thermocouple is mounted close to the sample. To determine the temperature difference between this thermocouple and the inside of the sample cell we preformed calibration measurements in which the solution was replaced by a second thermocouple, see Extended Data. Temperature scans were made using an Oxford Instruments ITC4 temperature controller. The time dependence of the temperature was measured using a temperature logger (TC-08 Thermocouple data logger, PicoLog) during all measurements. Graphs of the time-dependent temperatures are included in the Extended Data. The infrared spectra were recorded approximately every 18 seconds using a Perkin Spectrum Two FTIR spectrometer (spectral resolution 1 $cm^{-1}$).

# EXTENDED DATA

## Liquid-liquid transition in supercooled aqueous solution involving a low-temperature phase similar to low-density amorphous water

Sander Woutersen[*], Michiel Hilbers[*], Zuofeng Zhao[#] and C. Austen Angell[#]

[*]Van 't Hoff Institute for Molecular Sciences, University of Amsterdam, Science Park 904, 1098 XH Amsterdam, The Netherlands

[#]School of Molecular Sciences, Arizona State University, Tempe, AZ, 85287-1604, USA

**Extended Data**

*IR spectrum of NaTFA solution in HDO:D$_2$O*

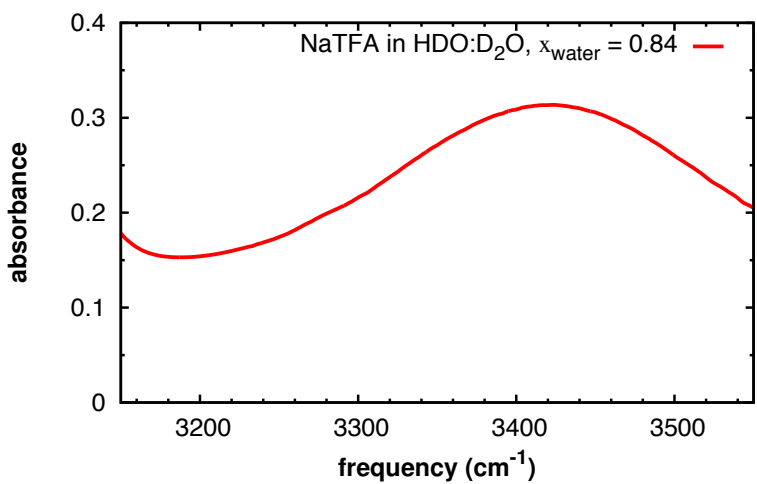

The shoulder at ~3300 cm$^{-1}$ observed in Fig. 2a,d of the main text for (N$_2$HD$_4$/N$_2$D$_5$)TFA in HDO:D$_2$O is lacking in the spectrum of NaTFA in HDO:D$_2$O ($x_{water}$ = 0.84). This shoulder can therefore be assigned to (NH-stretching of) N$_2$HD$_4^+$.

*Time dependence of the temperature during the measurements*

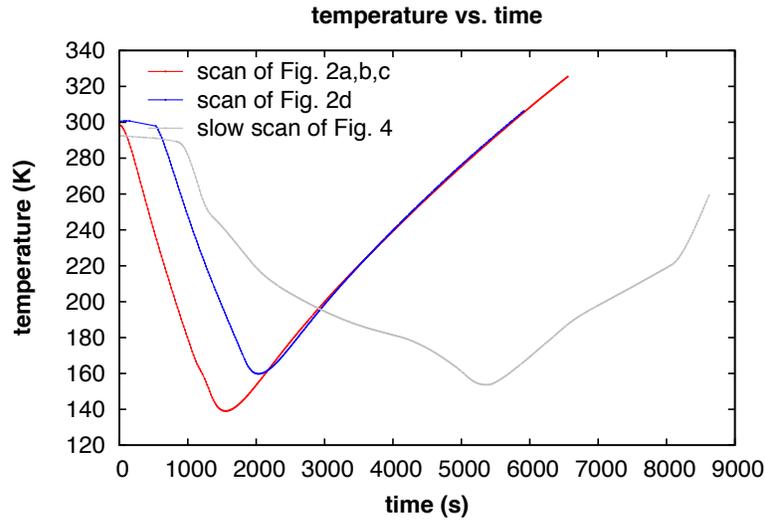

We cooled down and reheated the liquid samples at different cooling rates. Shown is the temperature of the sample vs. time for each of the measurements shown in the main text. Note the identical cooling rates (negative slope) during the measurements of Figs. 2a,b,c and d, and the slow cooling rate (1K/min) at temperatures around 200K in the measurement of Fig 4b. In all measurements, we cool the liquid sample down to a temperature well below the liquid-liquid transition temperature and subsequently reheat the liquid sample to confirm that the transition is reversible.

*Temperature calibration and temperature hysteresis*

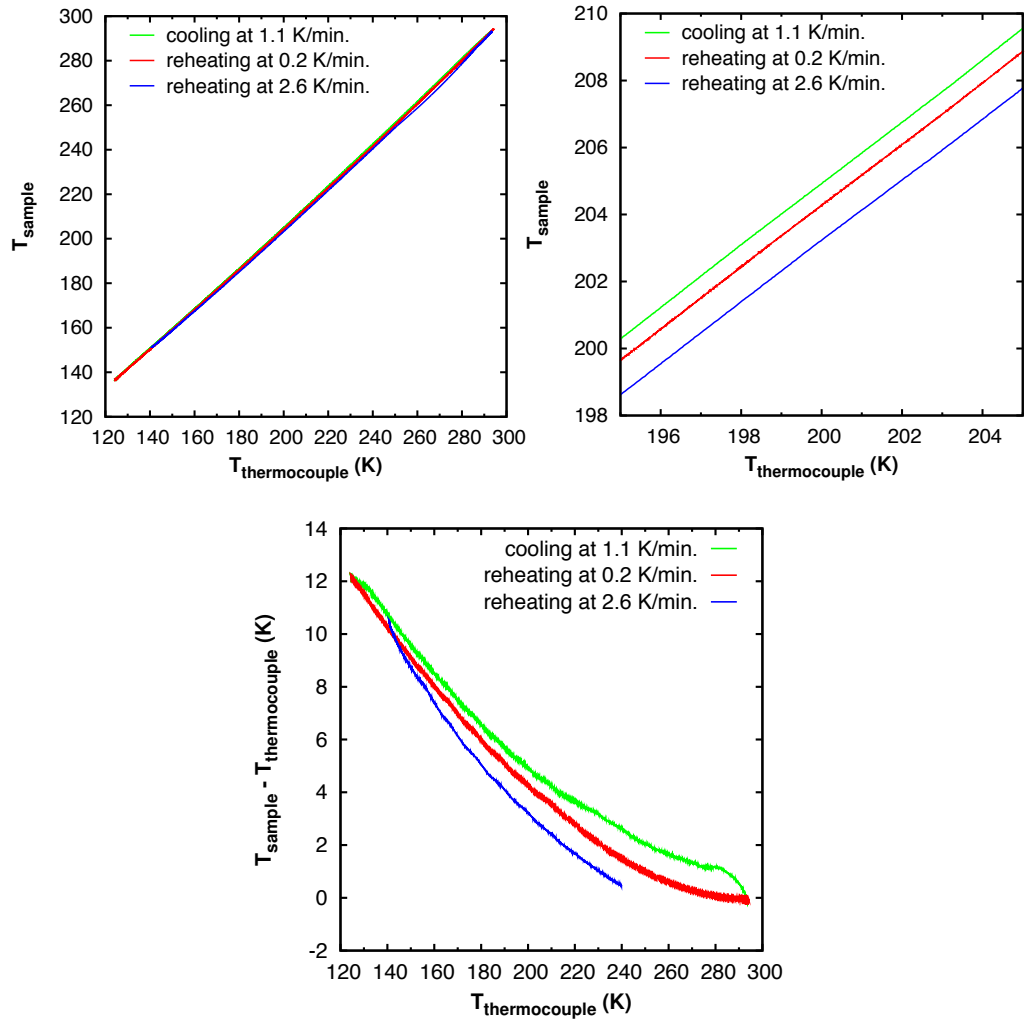

Upper two panels: in a separate set of temperature scans, we replaced the liquid sample by a second thermocouple (using thermal paste to ensure good thermal contact between the thermocouple and the sample-cell windows) in order to measure the temperature difference between the probe-thermocouple readout as used in the IR measurements and the actual temperature inside the (now empty) liquid-sample cell. We find that for temperature scanning rates comparable to the ones used in the IR measurements of Figs. 2 and 4, at 200 K the temperature in the cell is 4K higher than at the probe thermocouple. Lower panel: the temperature difference shows a hysteresis of ~2 K at temperatures around 200 K.